\documentclass{vgtc}

\ifpdf
  \pdfoutput=1\relax                   
  \usepackage{graphicx}                
  \DeclareGraphicsExtensions{.pdf,.png,.jpg,.jpeg} 
\else
  \usepackage{graphicx}                
  \DeclareGraphicsExtensions{.eps}     
\fi%

\graphicspath{{figures/}{pictures/}{images/}{./}} 

\usepackage{microtype}                 
\PassOptionsToPackage{warn}{textcomp}  
\usepackage{textcomp}                  
\usepackage{mathptmx}                  
\usepackage{times}                     
\usepackage{cite}                      
\usepackage{tabu}                      
\usepackage{booktabs}                  

\usepackage{lineno}

\usepackage[table]{xcolor}

\usepackage[caption=false,font=footnotesize,labelfont=sf,textfont=sf]{subfig}
\usepackage{multirow}

\usepackage[printonlyused,nolist]{acronym}
\acrodef{FFM}{Five-Factor Model}
\acrodef{GUI}{graphical user interface}
\acrodef{HCI}{Human-Computer Interaction}
\acrodef{IQR}{interquartile range}
\acrodef{LMLoC}{Levenson Multidimensional Locus of Control}
\acrodef{LoC}{Locus of Control}
\acrodef{MBTI}{Myers-Briggs Type Indicator}
\acrodef{NEO PI-R}{Revised NEO Personality Inventory}
\acrodef{UI}{user interface}
\acrodef{UX}{user experience}

\onlineid{1204}

\vgtccategory{Research}

\vgtcinsertpkg

\title{Exploring How Personality Models Information Visualization Preferences}

\author{Tomás Alves\thanks{e-mail: tomas.alves@tecnico.ulisboa.pt}, Bárbara Ramalho\thanks{e-mail: barbara.ramalho@tecnico.ulisboa.pt}, Daniel Gonçalves\thanks{e-mail: daniel.goncalves@inesc-id.pt}, and Sandra Gama\thanks{e-mail: sandra.gama@tecnico.ulisboa.pt}\\ %
        \scriptsize INESC-ID and Instituto Superior Técnico, \\ \scriptsize University of Lisbon, Lisbon, Portugal %
\and Joana Henriques-Calado\thanks{e-mail: joanahenriquescalado@gmail.com}\\ %
     \scriptsize CICPSI, Faculdade de Psicologia, \\
     \scriptsize Universidade de Lisboa, Lisboa, Portugal 
}

\teaser{
  \centering
  \includegraphics[width=.8\linewidth]{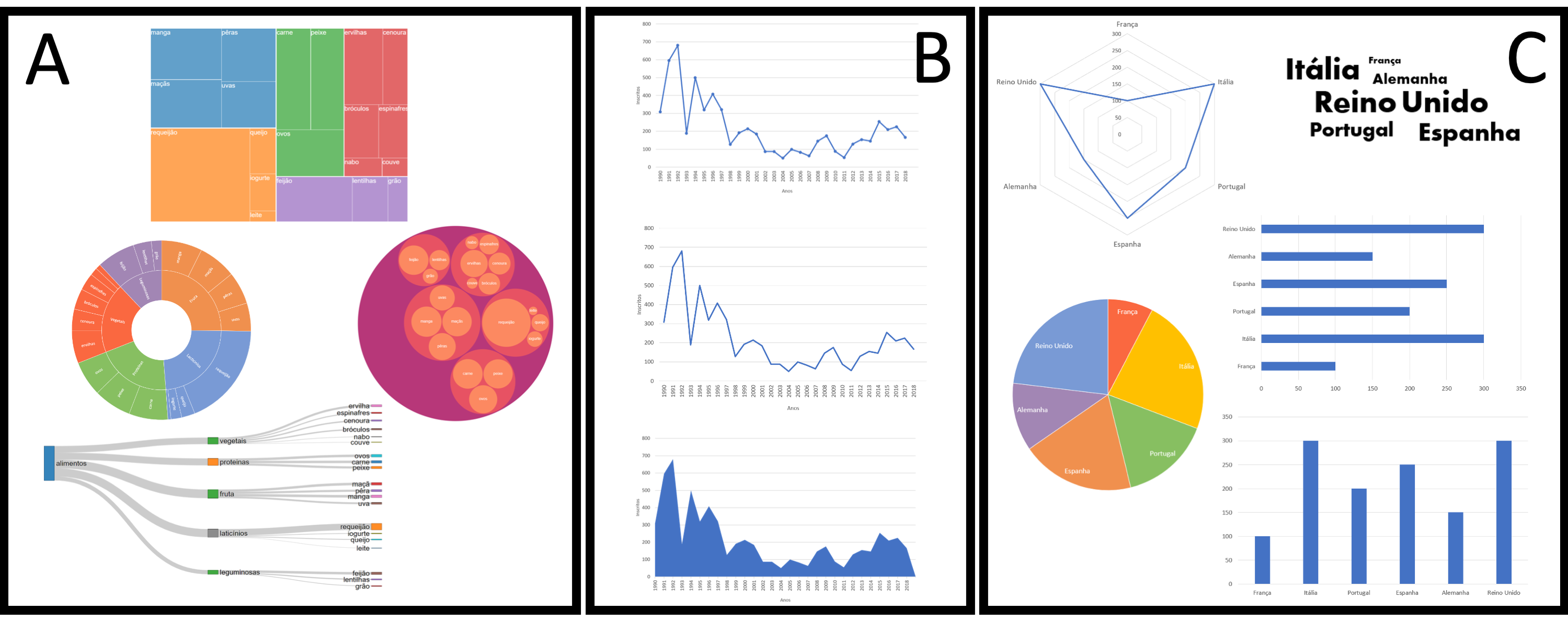}
  \caption{The three different information visualization contexts that we address in our study. \textbf{(A)} the hierarchy context which is studied with the treemap, circular packing, sunburst, and Sankey diagram idioms. \textbf{(B)} evolution \textcolor{black}{over time} context, where we apply line charts with and without points, and an area chart. Finally, \textbf{(C)} \textcolor{black}{comparison} context, which includes the radar chart, word cloud, horizontal and vertical bar charts, and pie chart idioms.}
  \label{fig:teaser}
}

\abstract{Recent research on information visualization has shown how individual differences act as a mediator on how users interact with visualization systems.
We focus our exploratory study on whether personality has an effect on user preferences regarding idioms used for hierarchy, evolution \textcolor{black}{over time}, and \textcolor{black}{comparison} contexts.
Specifically, we leverage all personality variables from the Five-Factor Model and the three dimensions from Locus of Control (LoC) with correlation and clustering approaches.
The correlation-based method \textcolor{black}{suggested} that Neuroticism, Openness to Experience, Agreeableness, several facets from each trait, and the External dimensions from LoC mediate how much individuals prefer certain idioms.
In addition, our results from the cluster-based analysis showed that Neuroticism, Extraversion, Conscientiousness, and all dimensions from LoC have an effect on preferences for idioms in hierarchy and evolution contexts.
Our results support the incorporation of in-depth personality synergies \textcolor{black}{with InfoVis} into the design pipeline of visualization systems.
} 

\CCScatlist{
  \CCScatTwelve{Human-centered computing}{Human computer interaction (HCI)}{HCI design and evaluation methods}{User studies};
  \CCScatTwelve{Human-centered computing}{Visu\-al\-iza\-tion}{Visualization design and evaluation methods}{}
}

\begin{document}


\firstsection{Introduction}

\maketitle

Individual differences have shown promise as an adaptation metric of information visualization systems to tackle the limitations of one-size-fits-all approaches~\cite{conati2015towards,oscar2017towards,lalle2019role}.
The inclusion of these factors empowers developers with guidelines on how individual characteristics impact human-computer interaction.
Among several psychological constructs that differentiate individuals such as cognitive bias or abilities, personality stands as \textcolor{black}{an} established strong mediator given its stability throughout adulthood~\cite{mccrae1982self}.
\textcolor{black}{Compared to other well studied individual cognitive traits (e.g. spatial ability and visual working memory), promising results regarding the relationship between personality traits and information visualization are few~\cite{liu2020survey}.
In order to bridge this gap, we focus on two of the most extensively studied personality models in our research field: the \ac{FFM}~\cite{costa1992neo} and the \ac{LoC}~\cite{lefcourt2014locus}.}

Although performance metrics such as speed or accuracy are important to address while users perform tasks in information visualization systems (e.g. ~\cite{ottley2015manipulating,ottley2015personality}), we believe that there is a lack of findings regarding how personality affects user preferences for information visualization techniques.
Weighting how personality has an effect on user preferences, our study focuses on whether \textbf{personality promotes user preferences regarding idioms used for hierarchy, evolution \textcolor{black}{over time}, and \textcolor{black}{comparison} contexts}, as these contexts have been frequently applied in state-of-the-art research~\cite{cashman2019inferential,liu2020survey,sheidin2020effect}.
In particular, we address all traits and their facets from the \ac{FFM}, and the three dimensions from \ac{LoC} to provide an in-depth analysis.
We take two distinct approaches to study this relationship: (i) correlation-based analysis -- where we investigate whether a personality variable had correlations with the user preference regarding an idiom -- and (ii) cluster-based analysis -- where we aggregate individuals by common characteristics and extract preference patterns from each group.
Our preliminary results \textcolor{black}{suggest} that personality has an effect on user preferences with both types of analysis.

\section{Related Work}
\label{sec:sota}

\begin{table*}
    \caption{\textcolor{black}{Traits and their facets of the Five-Factor Model.}}
    \label{tab:ffm}
    \centering
    \resizebox{.85\textwidth}{!}{%
    \begin{tabular}{ll}
        \hline\noalign{\smallskip}
        \textbf{Trait} & \textbf{Facets} \\
        \noalign{\smallskip}\hline\noalign{\smallskip}
        Neuroticism (N) & Anxiety (N1), Anger (N2), Depression (N3), Self-consciousness (N4), Immoderation (N5), Vulnerability (N6) \\ \rowcolor{lightgray}
        
        Extraversion (E) & Friendliness (E1), Gregariousness (E2), Assertiveness (E3), Activity level (E4), Excitement-seeking (E5), Cheerfulness (E6) \\
        
        Openness to Experience (O) & Imagination (O1), Artistic interests (O2), Emotionality (O3), Adventurousness (O4), Intellect (O5), Liberalism (O6) \\ \rowcolor{lightgray}
        
        Agreeableness (A) & Trust (A1), Morality (A2), Altruism (A3), Cooperation (A4), Modesty (A5), Sympathy (A6) \\
        
        Conscientiousness (C) & Self-efficacy (C1), Orderliness (C2), Dutifulness (C3), Achievement-striving (C4), Self-discipline (C5), Cautiousness (C6) \\
        \noalign{\smallskip}\hline
    \end{tabular}
    }
\end{table*}

\textcolor{black}{The Five-Factor Model consists of five general dimensions to describe personality and 30 subdimensions (facets) (Table~\ref{tab:ffm}): (i) Neuroticism distinguishes ``the stability of emotions and even-temperedness from negative emotionality"~\cite{Halko2010}; (ii) Extraversion suggests ``a lively approach toward the social and material world"~\cite{Halko2010}; (iii) Openness to Experience describes ``the wholeness and complexity of an individual's psychological and experiential life"~\cite{Halko2010}; (iv) Agreeableness distinguishes ``pro-social and communal orientation toward others from antagonism"~\cite{Halko2010}; and (v) Conscientiousness suggests ``self-use of socially prescribed restraints that facilitate goal completion, following norms and rules, and prioritizing tasks"~\cite{Halko2010}.}
Ziemkiewicz and Kosara~\cite{ziemkiewicz2009preconceptions} found that high Openness to Experience led individuals to be faster while solving problems related to hierarchical visualizations that include conflicting visual and verbal metaphors.
Furthermore, Ziemkiewicz et al.~\cite{ziemkiewicz2012visualization} concluded that neurotic individuals attained high accuracy on hierarchical search tasks.
Introverted participants were more accurate in answering the questions posed by the tasks.
Other contributions~\cite{green2010towards,ziemkiewicz2011locus,brown2014finding,braun2019user,oscar2017towards} have addressed the traits of Neuroticism and Extraversion.
\textcolor{black}{Results have shown how these traits have an effect on task performance metrics such as the time to complete a task~\cite{green2010towards,oscar2017towards} and accuracy~\cite{oscar2017towards}.
Additionally, Neuroticism and Openness to Experience exhibited an effect on the attractiveness and dependability ratings from participants regarding driver state visualization systems~\cite{braun2019user}.}


The Locus of Control orientations are described as two different aspects, which are distinguished by different reinforcements.
While internal \ac{LoC} is related with internal reinforcement because the value of an individual is heightened by some event or environment, external \ac{LoC} is linked to external reinforcement since it addresses how some event or environment yields benefit for the group or culture to which the individual belongs to~\cite{kim2016motivating}.
Furthermore, external \ac{LoC} can be differentiated in two types: \textit{Powerful Others} -- believe in an ordered world controlled by powerful others -- and \textit{Chance} -- consider the world as unordered and chaotic~\cite{levenson1973reliability}.
Several studies have shown how \ac{LoC} is related to search performance across hierarchical~\cite{green2010towards}, \textcolor{black}{time series~\cite{sheidin2020effect}, and item comparison~\cite{cashman2019inferential} visualization designs}, visualization use~\cite{ziemkiewicz2011locus,ziemkiewicz2012visualization}, and behavioural patterns~\cite{ottley2015personality}.
Although \textit{Internals} are significantly faster than \textit{Externals} when performing procedural tasks (search tasks to locate items)~\cite{green2010towards}, \textit{Externals} are faster and more accurate than \textit{Internals} regarding inferential tasks such as comparing two items~\cite{ziemkiewicz2011locus,ziemkiewicz2012visualization}.
In addition, \textit{Internals} are usually faster than \textit{Externals} in image-based search tasks~\cite{brown2014finding}.

Although performance metrics such as speed or accuracy are important to address, there are strong results regarding user preferences in information visualization.
Ziemkiewicz et al.~\cite{ziemkiewicz2012visualization} focus on Neuroticism, Extraversion, and the \ac{LoC}, while Lallé and Conati~\cite{lalle2019role} address the latter.
In contrast, Toker et al.~\cite{toker2012towards} did not address personality.
Nevertheless, research has not found effects for Agreeableness or Conscientiousness~\cite{liu2020survey}, and \ac{FFM} traits' facets have been neglected.
In our study, we propose an extension of the state-of-the-art research by including the remaining \ac{FFM} traits and their facets, which may hinder relationships that are only represented at a finer granularity of personality variables.

\section{Data Collection}
\label{sec:method}

In order to study \textbf{\emph{how personality affects user preferences regarding information visualization techniques}}, we started by choosing which \textcolor{black}{contexts we wanted to address (Figure~\ref{fig:teaser}): (i) hierarchy, one of the most common in research (e.g.~\cite{ziemkiewicz2012visualization}); (ii) evolution over time, giving the importance of time series data analysis~\cite{sheidin2020effect}; and (iii) comparison, as it is more appropriate to show differences or similarities between values at a fixed granularity~\cite{cashman2019inferential}.}
We include a simple and familiar scenario with each context in order to stimulate users to reflect on the implications of using each idiom rather than the complexity of the data.
We focused on minimizing the number of channels and marks of each graph and keeping them consistent across contexts, while keeping the same data within a context.

Regarding hierarchy, items are all related to each other by the principle of containment.
We opted for a treemap, a circular packing diagram, a sunburst, and a Sankey diagram to display the distribution of food consumed by a household within a month.
For evolution \textcolor{black}{over time} contexts, we chose line charts with and without points, and area charts.
The scenario asked the participant to imagine that the data referred to the number of registrants and participants in a marathon held annually in the United States.
Finally, we decided to use radar charts, word clouds, horizontal and vertical bar charts, and pie charts for the \textcolor{black}{comparison} context.
In particular, the scenario represents the levels of the happiness index among six different countries (France, Italy, Portugal, Spain, Germany, and the United Kingdom).


Participants were recruited through standard convenience sampling procedures including direct contact and through word of mouth.
Our final data set comprises 64 participants (30 males, 34 females) between 18 and 60 years old \((M = 24.27; SD = 7.10)\).
\textcolor{black}{In addition, they were asked whether they were using glasses or contact lenses and the apparatus used while filling in the questionnaire.
Neither factor had a significant effect on the experience.}

\textcolor{black}{Before the experiment, participants were informed about the experience and invited to agree with a compulsory consent form.
They were also informed that they could quit the experiment at any time.
We then collected the \ac{FFM} five personality traits and its 30 facets, and the dimensions of \ac{LoC} with the Portuguese versions of the \ac{NEO PI-R}~\cite{costa2008revised,lima2000neo} and the IPC scale~\cite{levenson1973multidimensional,queiros2004burnout}, respectively.
Afterwards, participants were presented an online questionnaire which contained a visual example of each idiom grouped by context.
Participants were firstly prompt to read the scenario for the respective context and then assess their preference for an idiom by completing a seven-point Likert scale ranging from \textit{Low Preference} (1) to \textit{High Preference} (7).
We allowed participants to freely change their ratings until they were satisfied with all ratings in order to avoid the anchoring bias.
}

\section{Correlation-based Analysis}

In order to find correlations between personality variables and user preferences, we used the Spearman's correlation method (Table~\ref{tbl:spearman}), as it is preferable when variables feature heavy-tailed distributions or when outliers are present~\cite{de2016comparing}, and it has been shown as an appropriate statistical analysis with Likert scales~\cite{norman2010likert}.
\textcolor{black}{Our hypothesis is that a personality dimension from the \ac{FFM} and/or the \ac{LoC} is correlated with how participants rated their preference for an idiom.
Taking into account the large number of statistical models, we use a Bonferroni correction to counteract the problem of multiple comparisons.
Therefore, significant p-values are reported at $\alpha = 0.0001$.
Although we did not find any statistical significance, results suggest that, at} a trait level, Neuroticism, Openness to Experience, and Agreeableness show weak negative effects with line charts with points \((r\textsubscript{s}(64) = -.267, p = .033)\), area charts \((r\textsubscript{s}(64) = -.29, p = .02)\) and sunburst \((r\textsubscript{s}(64) = -.285, p = .022)\), respectively.
In addition, we found that 19 facets showed similarly weak effects.
Among these facets, we can observe that facets from Agreeableness \textcolor{black}{pointed towards} more effects, followed by Neuroticism and Openness to Experience.
Although both Extraversion and Conscientiousness did not have an effect, three facets from each of these traits \textcolor{black}{imply an} effect.
Regarding \ac{LoC}, both External dimensions showed weak positive correlations.
While Powerful Others \textcolor{black}{may have} modelled how participants rated both area \((r\textsubscript{s}(64) = .32, p = .01)\) and line chart without points \((r\textsubscript{s}(64) = .313, p = .012)\), Chance \textcolor{black}{hinted} an effect on ratings for the pie chart \((r\textsubscript{s}(64) = .382, p = .002)\).
Taking into account the idioms, we can see that line charts \textcolor{black}{suggest} the largest number of effects related to personality-based user preferences, as most of Neuroticism, its facets, and facets from Conscientiousness \textcolor{black}{suggested} correlation effects.
At a broader level, evolution \textcolor{black}{over time} context idioms \textcolor{black}{indicated} the largest number of effects (53.57\%), followed by hierarchy (28.57\%) and then \textcolor{black}{comparison} (17.86\%).

\begin{table}
    \caption{Significant results from the Spearman's correlation tests.}
    \label{tbl:spearman}
    \centering
    \resizebox{.75\columnwidth}{!}{%
    \begin{tabular}{ccrc}
        \hline\noalign{\smallskip}
        \textbf{Personality} & \textbf{Idiom} & \multicolumn{1}{c}{\textbf{$r\textsubscript{s}$}} & \multicolumn{1}{c}{\textbf{p-value}} \\
        \noalign{\smallskip}\hline\noalign{\smallskip}
        N & Line Chart with Points & -0.267 & 0.033 \\
        N1 & Treemap & -0.364 & 0.003 \\
        N3 & Line Chart with Points & -0.292 & 0.019 \\
        N4 & Line Chart with Points & -0.276 & 0.027 \\
        N6 & Line Chart with Points & -0.277 & 0.027 \\
        E1 & Sunburst & -0.274 & 0.029 \\
        E4 & Line Chart with Points & 0.247 & 0.049 \\
        E6 & Line Chart without Points & -0.283 & 0.024 \\
        O & Area Chart & -0.290 & 0.020 \\
        O1 & Horizontal Bar Chart & -0.269 & 0.032 \\
        O2 & Line Chart without Points & -0.251 & 0.046 \\
        O3 & Area Chart & -0.320 & 0.010 \\
        O5 & Area Chart & -0.340 & 0.006 \\
        O6 & Sankey Diagram & -0.268 & 0.032 \\
        A & Sunburst & -0.285 & 0.022 \\
        A2 & Sunburst & -0.317 & 0.011 \\
        A2 & Treemap & -0.275 & 0.028 \\
        A3 & Line Chart without Points & -0.249 & 0.047 \\
        A5 & Radar Chart & -0.312 & 0.012 \\
        A5 & Circular Packing & 0.249 & 0.047 \\
        A6 & Sunburst & -0.277 & 0.027 \\
        A6 & Radar Chart & -0.263 & 0.036 \\
        C3 & Line Chart with Points & 0.255 & 0.042 \\
        C5 & Line Chart without Points & -0.246 & 0.050 \\
        C6 & Vertical Bar Chart & 0.274 & 0.028 \\
        Powerful Others & Area Chart & 0.320 & 0.010 \\
        Powerful Others & Line Chart without Points & 0.313 & 0.012 \\
        Chance & Pie Chart & 0.382 & 0.002 \\
        \noalign{\smallskip}\hline
    \end{tabular}
    }
\end{table}

\section{Cluster-based Analysis}

Following the work of Sarsam and Al-Samarraie~\cite{Sarsam2018}, we applied hierarchical density based clustering~\cite{han2011data,mcinnes2017hdbscan} to find that the most appropriate number of clusters to work with was three through silhouette and Davies–Bouldin index scores analysis~\cite{petrovic2006comparison} and Ward’s cluster method. Then, we used the k-means clustering algorithm~\cite{wang2015fast} to avoid the noise labels that hierarchical density based clustering yields.
We started by normalizing our data and allow the algorithm to run 100 iterations with different centroid seeds using Euclidean distance.
The final result contained the best output of 100 consecutive runs in terms of inertia.
As a follow-up, we conducted an ANOVA to validate whether each personality trait from one cluster differs from the other instances in the other clusters.
We found a significant difference $(p < .05)$ in between the three clusters regarding Neuroticism, Extraversion, Concientiousness, all dimensions from Locus of Control, and 18 personality facets out of 30.
These results show that all clusters have participants that differ among themselves in the aforementioned personality variables.
Table~\ref{tab:clusters} depicts the means and standard deviation values for all personality traits of the \ac{FFM} and dimensions from the \ac{LoC}.
The first cluster $(N = 35)$ notably has participants with the \textbf{highest levels of Conscientiousness and Internal dimension} across clusters.
It also includes people with the \textbf{lowest values on Neuroticism and the External dimensions}.
The remaining traits of Extraversion and Agreeableness show medium values, while Openness to Experience presents low levels.
In contrast, the second cluster $(N = 11)$ shows the \textbf{lowest values for Conscientiousness}.
In addition, it features participants with the \textbf{highest values of Extraversion and Agreeableness}, while the remaining personality variables show medium values among the clusters.
Finally, the third cluster $(N = 18)$ includes participants with the \textbf{highest levels on Neuroticism and on both the External dimensions}.
Nevertheless, it presents medium values for Conscientiousness and the remaining variables have each the lowest values of the set.
As we mentioned, it is possible to observe that the trait of Agreeableness presents similar values across clusters, while Openness to Experience, in spite of not showing significant differences between clusters, has very dissimilar values on Cluster 2 compared to the others.
In the real world, Cluster 1 contains people that are organized and believe in their efforts.
In contrast, Cluster 3 includes moody people that believe the external world has a large influence over their life.
Finally, Cluster 2 contains outgoing and open people.

\begin{table}
    \caption{Results of the K-means clustering algorithm for each personality trait and dimension.}
    \label{tab:clusters}
    \centering
    \resizebox{\columnwidth}{!}{%
    \begin{tabular}{lrrrrrr}
        \hline\noalign{\smallskip}
        \multirow{2}{*}{\textbf{Personality Variable}} & \multicolumn{2}{c}{\textbf{Cluster 1}} & \multicolumn{2}{c}{\textbf{Cluster 2}} & \multicolumn{2}{c}{\textbf{Cluster 3}} \\ 
        & \multicolumn{1}{c}{\textbf{M}} & \multicolumn{1}{c}{\textbf{SD}} & \multicolumn{1}{c}{\textbf{M}} & \multicolumn{1}{c}{\textbf{SD}} & \multicolumn{1}{c}{\textbf{M}} & \multicolumn{1}{c}{\textbf{SD}} \\
        \noalign{\smallskip}\hline\noalign{\smallskip}
        Neuroticism & 84.23 & 20.23 & 97.18 & 16.54 & \textbf{124.22} & \textbf{18.10} \\
        Extraversion & 112.00 & 19.02 & \textbf{122.00} & \textbf{13.82} & 92.94 & 16.11 \\
        Openness to Experience & 121.34 & 19.43 & \textbf{136.09} & \textbf{23.70} & 121.89 & 15.91 \\
        Agreeableness & 125.40 & 15.70 & 129.09 & 21.42 & 123.67 & 16.84 \\
        Conscientiousness & \textbf{140.74} & \textbf{15.85} & 96.73 & 20.03 & 105.67 & 18.24 \\
        Internal & \textbf{33.00} & \textbf{5.89} & 32.82 & 4.31 & 29.33 & 6.00 \\
        Powerful Others & 15.31 & 6.54 & 16.45 & 6.64 & \textbf{19.39} & \textbf{5.25} \\
        Chance & 15.97 & 5.87 & 19.82 & 5.84 & \textbf{20.17} & \textbf{5.84} \\
        \noalign{\smallskip}\hline
    \end{tabular}%
    }
\end{table}

In order to extract information visualization preferences for the different contexts among individuals of those three clusters, we opted for the Apriori algorithm~\cite{ilayaraja2013mining}, an association rules method to find common patterns.
Data preprocessing included the creation of an array for each participant containing the idiom that they preferred the most for each context.
In case of a tie between two or more idioms in their preference ratings, we included all idioms that tied together.
Afterwards, we divided users by their cluster labels and used the Apriori algorithm in each cluster.
Each run was performed with lower bound minimal values of 0.1 for support, 0.8 for confidence, and 3.1 for lift.
An Apriori association rule is often represented as $itemA \rightarrow itemB$, which translates into $itemB$ being frequently present in a set of preferences that also contains $itemA$.

\begin{table}
    \caption{Context and preferred idioms \textcolor{black}{with their frequency on top rules} for each cluster.}
    \label{tab:clusterIdioms}
    \centering
    \resizebox{\columnwidth}{!}{%
    \begin{tabular}{llll}
        \hline\noalign{\smallskip}
        \textbf{Context} & \textbf{Cluster 1} & \textbf{Cluster 2} & \textbf{Cluster 3} \\
        \noalign{\smallskip}\hline\noalign{\smallskip}
        \textit{Hierarchy} & Sunburst \textcolor{black}{(50\%)} & Sunburst \textcolor{black}{(76\%)} & Treemap \textcolor{black}{(37\%)} \\
        \textit{Evolution} & Line Chart w/ Points \textcolor{black}{(100\%)} & Line Chart w/out Points \textcolor{black}{(70\%)} & Line Chart w/out Points \textcolor{black}{(74\%)} \\
        \textit{\textcolor{black}{Comparison}} & Horizontal Bar Chart \textcolor{black}{(71\%)} & Horizontal Bar Chart \textcolor{black}{(71\%)} & Horizontal Bar Chart \textcolor{black}{(71\%)} \\
        \noalign{\smallskip}\hline
    \end{tabular}%
    }
\end{table}

We continued our analysis by choosing which rules to focus on the information visualization techniques according to the frequency of each rule.
We started by choosing the rule with the highest frequency and then choosing rules that had similar item sets until there was no rule with common or contradictory associations.
Finally, if a context did not have a style associated to it, we chose the most frequent preferred idiom for that context among participants of the cluster, which was the case for the hierarchy context for Cluster 1.
Based on the final set of rules for each cluster, we were able to derive which idioms were the most preferred according to the different contexts (Table~\ref{tab:clusterIdioms}).
Notably, there are differences in the contexts of evolution \textcolor{black}{over time} and hierarchy.
Regarding the evolution \textcolor{black}{over time} context, both Clusters 1 and 2 prefer a sunburst idiom and Cluster 3 participants rate treemaps higher.
\textcolor{black}{Compared to the other contexts, the chosen evolution over time idiom was less prominent in Clusters 1 (50\%) and 3 (36.7\%), while Cluster 2 was more consistent in their preference (76.2\%).
For hierarchy contexts, while Cluster 1 completely prefers line charts with points (100\%), the remaining clusters would rather omit the use of those marks.
Finally, all clusters state that an horizontal bar chart is the most preferred idiom to use for comparison data, with frequency values around 71\%.
This may hint that the appropriateness of an idiom for a specific problem context acts as a stronger regulator compared to personality.}

\section{Discussion}

After analysing our results with both approaches we were able to have a better understanding of how personality has an effect on information visualization technique preferences.
From the correlation-based analysis, \textcolor{black}{results pointed towards effects from the Neuroticism, Openness to Experience, and Agreeableness traits in user preference regarding different idioms.
Several more facets from all \ac{FFM} traits and both External \ac{LoC} dimensions also suggested a correlation.
This lack of significance results is a consequence of the Bonferroni correction we applied in order to counteract the multiple comparisons problem.
We believe the correlation-based approach is sound for a smaller number of Spearman correlations, which points our next step in this research towards the separate analysis of these personality variables to verify whether the results of this study have a high false discovery rate.}
Regarding Neuroticism, the trait itself and three facets showed a weak negative effect, suggesting that people with higher levels of Neuroticism dislike line charts with points.
In fact, we were able to verify the same effect on Cluster 3, where users had the highest levels of Neuroticism and they preferred line charts without points.
Additionally, only the cluster with the lowest levels of Neuroticism (Cluster 1) showed a preference towards line charts with points.
This effect may be given to how people with high Neuroticism experience more stress when the idiom contains more marks, thus more information that may be harder to perceive.
Extraversion only showed strong results in the cluster-based approach.
While individuals with high and medium levels preferred sunburst as an idiom to represent hierarchy, \textcolor{black}{people with} low levels showed a preference towards treemaps.
Interestingly, individuals with medium levels would rather use line charts with points, contrary to the remaining participants which showed an inclination towards excluding points on those charts.
This effect may be explained by interaction effects with the remaining personality variables.
In the correlation-based approach, three of its facets \textcolor{black}{may have had an effect on} preferences for sunburst and line chart idioms, yet all effects were weak in size.

The best clusters produced by the k-means algorithm did not divide individuals significantly based on Openness to Experience.
Nevertheless, we found that independently of the remaining personality variables, \textcolor{black}{results suggested it} could foster the preference for area charts.
Furthermore, five of its facets \textcolor{black}{hinted} negative weak effects with several idioms, mostly regarding evolution \textcolor{black}{over time} idioms.
This is a rather interesting effect, considering how Openness to Experience has been shown to model how individuals process evolution~\cite{feist2004openness,acerbi2009cultural}.
Agreeableness was also not significantly different among the different clusters, yet, similarly to Openness to Experience, it \textcolor{black}{suggested} some significant effects on the correlation-based approach along four of its facets.
Most of these effects are referring to hierarchy, which may be related to how Agreeableness models how people evaluate hierarchical structures of collectivism~\cite{realo1997hierarchical}.
\textcolor{black}{Finally, Conscientiousness} showed more effects while interacting with the other personality variables in the cluster-based approach then by an analysis with correlations.
People with high levels tend to prefer line charts with points compared to the remaining population.
We believe that this preference may be given to how these people prefer an organised approach to life, thus preferring to see idioms with more detail.
We also found that people with high and low values on this trait prefer a sunburst in comparison to a treemap, similar to the Extraversion trait.
Concerning the dimensions from \ac{LoC}, both External dimensions \textcolor{black}{suggested} positive weak correlation effects.
While Powerful Others \textcolor{black}{hinted} an effect on the evolution \textcolor{black}{over time} context, Chance \textcolor{black}{did it} for the \textcolor{black}{comparison} context.
In addition, cluster-based analysis showed that people with higher values on these dimensions and the lowest Internal levels among the clusters have a preference for a treemap compared to the sunburst idiom.
In contrast, the highest values for Internal and lowest for both the External dimensions showed a preference for line charts with points.
This effect may be a result of \textit{Internals} being faster than \textit{Externals} when the former search for items~\cite{green2010towards} because they use additional marks such as points to guide their search.

\textcolor{black}{In the light of this, our results suggest} that personality is a differentiating factor when it comes to designing information visualization systems.
\textcolor{black}{Looking into our approaches, for example, while facets from Conscientiousness hinted in the correlation-based approach, participants from Cluster 1 preferred line charts with points.
In addition, results from the Cluster 3 were indicated by the correlation results where higher values on Neuroticism or its facets led participants to choose a line chart without points.}
The same effect happened with Powerful Others.
In contrast, we found dissimilar effects for Extraversion and Agreeableness.
Regarding the former, we expected that Cluster 2 would have a preference on the evolution \textcolor{black}{over time} context for line charts with points and on the evolution \textcolor{black}{over time} context different from a sunburst idiom.
Moreover, the latter was not significantly different between clusters.
In the light of this, we hypothesize that this lack of significance led to an omission of interaction effects from Agreeableness.
In this case, we must also address how participants rated each idiom independently of personality.
As the Spearman's correlation effects were all small in size, we believe that the interaction effects were stronger, as one individual perceives information through interactions of all their personality constructs and not only one.
Thus, we consider the cluster-based approach to be more appropriated.
There are some important factors that may explain the lack of significance observed in some of our results.
First, since we are tackling a lot of personality variables, a larger number of participants would allow conclusions with a stronger impact in both approaches.
In particular, we could have a better sampling regarding Openness to Experience, Agreeableness, and the Internal dimension of \ac{LoC}.
Secondly, although there are more idioms in information visualization, \textcolor{black}{there are more idioms from these contexts to explore.
We also} did not control for the familiarity cognitive bias, which may have had an effect on the results.
Thirdly, the scenario \textcolor{black}{and the complexity of the dataset} used to illustrate the different contexts may have had an effect on how people perceived the idioms.
Finally, not asking users to perform any task rather than rating their preference for the aesthetics of an idiom may not impact visual task analysis.

\section{Conclusions and Future Work}

This exploratory study \textcolor{black}{focuses on} personality with two different psychological constructs (\ac{FFM} and \ac{LoC}) models user preferences regarding information visualization techniques in three different contexts: hierarchy, evolution \textcolor{black}{over time}, and \textcolor{black}{comparison}.
Besides identifying which idioms are modelled by personality-based user preferences, our results \textcolor{black}{suggest} important implications that may be used in the design pipeline to customize information visualization systems.
Future work includes the implementation and testing of different information visualization systems developed based on our results to assess how they affect user preference, performance, experience, and satisfaction.
In addition, task types, \textcolor{black}{task complexity},  and contexts should be further explored as they may lead to distinct interactions of users given their individual differences.
Finally, we aim to recruit a larger number of participants so that we can explore more in-depth the personality variables that were not significantly different between clusters.

\acknowledgments{
This work was supported by national funds through Fundação para a Ciência e a Tecnologia (FCT) with references UIDB/50021/2020 and SFRH/BD/144798/2019.}

\bibliographystyle{abbrv-doi}

\bibliography{template}
\end{document}